\documentclass[aps,prb,twocolumn,superscriptaddress]{revtex4-2}
\usepackage{bm}                     
\usepackage{graphicx}               
\usepackage{graphics}               
\usepackage{amsmath}                
\usepackage{amsfonts}               
\usepackage{amssymb}                
\usepackage{makeidx}                
\usepackage{wasysym}                
\usepackage{bbold}                  
\usepackage[makeroom]{cancel}       
\usepackage[utf8]{inputenc}         
\usepackage[normalem]{ulem}         
\usepackage{xcolor}                 
\usepackage{soul}                   
\usepackage{float}                  
\usepackage{physics}                
\usepackage{IEEEtrantools}          
\usepackage{mathrsfs} 

\usepackage[T1]{fontenc}
\usepackage[utf8]{inputenc}

\newcommand{\be}{\begin{equation}}
\newcommand{\ee}{\end{equation}}
\newcommand{\ba}   {\begin{eqnarray}}
\newcommand{\ea}   {\end{eqnarray}}

\usepackage[unicode=true,pdfusetitle,
 bookmarks=true,bookmarksnumbered=false,bookmarksopen=false,
 breaklinks=false,pdfborder={0 0 1},backref=false,colorlinks=false]
 {hyperref}
\hypersetup{
 colorlinks,linkcolor=darkorange,citecolor=magenta,urlcolor=royalblue}

\definecolor{navy}{RGB}{0,0,128}
\definecolor{royalblue}{RGB}{65,105,225}
\definecolor{darkorange}{RGB}{204,85,0} 
\definecolor{forestgreen}{RGB}{34,139,34}

\begin{document}

\title{Chiral-symmetry breaking and degeneracy in Koch fractal geometries}

\author{L. L. Lage}
\email{lucaslagedoc@gmail.com}
\affiliation{Instituto de F\'isica, Universidade Federal Fluminense, Niter\'oi, Av. Litor\^{a}nea sn 24210-340, RJ-Brazil}

\author{A. Latg\'e}
\affiliation{Instituto de F\'isica, Universidade Federal Fluminense, Niter\'oi, Av. Litor\^{a}nea sn 24210-340, RJ-Brazil}

\date{\today}

\begin{abstract}
Fractal geometries provide a distinctive platform for controlling quantum interference, topology, and localization. We investigate the Su--Schrieffer--Heeger (SSH)  model constructed on the Koch curve, where a triangular geometry enables additional same-sublattice hoppings that break chiral symmetry and modify the conventional topological behavior captured by a real-space local marker. Destructive interference within the triangular units produces a highly degenerate flat level at $E=-t$. Although this level originates from the local geometry, its degeneracy and spectral weight are controlled by the self-similar connectivity of the Koch curve. The hierarchy reduces the number of independent flat states and reorganizes the spectrum into an intricate distribution of energy levels. Our results reveal complementary roles for local triangular interference and global fractal connectivity in shaping the electronic properties of geometrically modified SSH chains.



\end{abstract}

\maketitle

\section{Introduction}

The Koch curve is a self-similar fractal obtained by dividing each line segment into three equal parts and replacing the middle segment with two sides of a triangle. Repeating this procedure generates triangular structures at progressively smaller length scales and gives the Hausdorff dimension $D_{\mathrm{K}}=\log(4)/\log(3) \approx 1.26$ \cite{VonKoch1904}. When used as a chain, this hierarchical geometry can produce spectral gaps, unconventional localization, and boundary states beyond those found in regular systems \cite{PaiPrem2019,CristianePRL,Biswas2023,lagefrontier,Salib2024}. Related effects have also been investigated in photonic and acoustic lattices, where the geometry and hopping strengths can be controlled experimentally  \cite{CaceresAravena2022,Song2026,Li2023Fractal,BlancoRedondo2016,Zheng2022AcousticFractal,Li2022AcousticFractalHOTI,Li2023}. Moreover, acoustic lattices may be constructed from coupled cavities, resonators, or waveguide channels, where the effective hopping strengths are tuned through channel widths, or intercavity distances \cite{Gao2023Acoustic,Coutant2021}. An important question is how to distinguish properties produced by the local triangular geometry from those caused by the global Koch hierarchy.

The Su--Schrieffer--Heeger (SSH) chain is a paradigmatic
one-dimensional symmetry-protected topological insulator. The combination of protected time-reversal and chiral symmetries defines the spectral particle-hole symmetry, placing the
conventional SSH model in the BDI symmetry class
\cite{Chiu2016,Altland1997}. In one dimension, this class admits an integer winding number, which distinguishes the trivial and nontrivial dimerization regimes. Moreover, chiral symmetry enforces the spectral electron-hole correspondence and produces boundary states within nontrivial phase at zero energy in such systems
\cite{SSH1979,Atala2013,Lu2024}. Extensions of this model to coupled
chains, and interacting regimes have revealed further phase
diagrams and different localization configurations
\cite{Mao2026,Nersesyan2020,PhysRevB.109.195124,
PhysRevB.98.024205,Xu2022,Melo2023,Tiao2026}. 

In systems lacking of translational invariance the use of real-space resolved local markers plays an important role to classify topology in such systems with different topological classes \cite{Weichen2023,CristianeHaldane,LageSOC,TitusSC2018,Bianco2011,Hannukainen2022,Oliveira2024}. Here, we investigate the SSH model defined on a Koch curve with triangular structures arranged at different hierarchical levels. Hoppings between consecutive sites preserve the bipartite SSH chain, while additional hoppings $t'$ across the triangle bases connect sites of the same sublattice and break chiral symmetry. Therefore the Koch--SSH model plays between the chiral-symmetric limit at $t'=0$ and a chiral-broken regime for $t'\neq0$. We find that $t'$ converts the SSH gap closing into a gapped transition between the two dimerization
regimes and continuously drives the local marker away from its
quantized values.

We further separate the local and hierarchical effects by comparing the Koch curve with a size-matched prolonged chain formed by repeated $G(1)$ units. Both systems contain a highly degenerate level at $E=-t$, which originates from destructive interference within the common triangular geometry. In fact, the Koch hierarchy introduces additional geometrical constraints and removes one third of the states at $E=-t$ also  found in the prolonged chain. Its spectral weight consequently approaches $1/6$ in the Koch fractal, compared with $1/4$ in the prolonged structure. The hierarchy also splits the spectrum into staircase-like clusters and produces spatially modulated states absent in a nonfractal structure. These results show that the local triangular geometry generates the flat states, while the intricate Koch connectivity controls its degeneracy, promotes topological symmetry breaking and reorganizes the electronic spectrum.


\section{Koch--SSH Model}
\begin{figure*}[t!]
\centering   
\includegraphics[width=\textwidth]{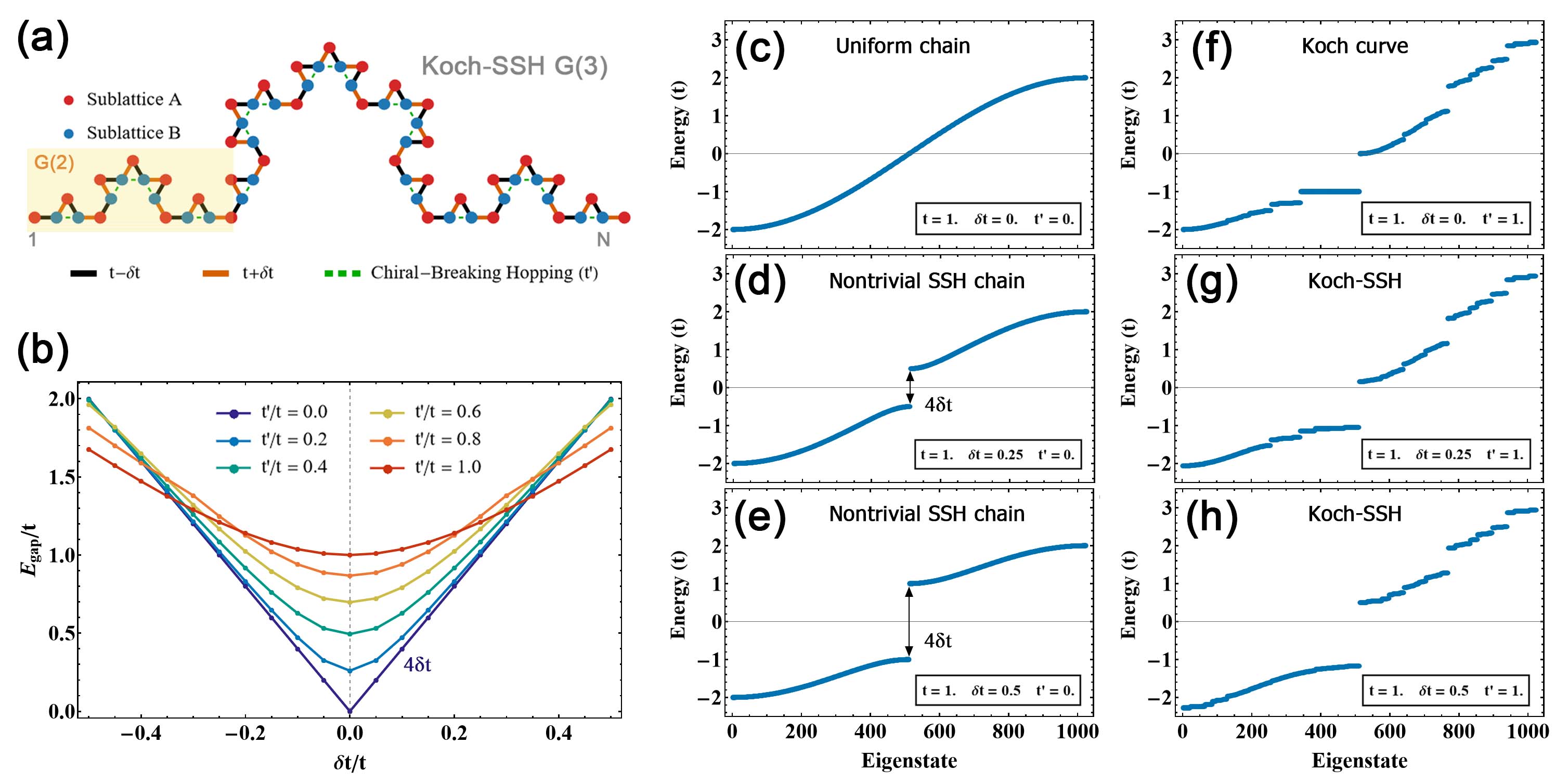}
\caption{(a) Hopping scheme of the Koch--SSH model in a G(3) fractal curve. Alternating hoppings $t-\delta t$ (black) and $t + \delta t$ (orange) preserves the bipartite SSH structure, while $t'$ (green dashed lines) connects sites belonging to same sublattice within the Koch triangles bases. (b) Energy gap $E_{\text{gap}}$ as a function of the dimerization $\delta t/ t$ and the chiral-breaking hopping $t'/t$ for a Koch--SSH G(5). Panels (c)--(h) show representative energy spectra of Koch--SSH G(5) for different combinations of energy hopping parameters.}
\label{FIG_CHIRALKOCH}
\end{figure*}
We consider an SSH model defined on a Koch fractal of $n$-th generation $G(n)$, containing $N$ sites at positions $\mathbf{r}_i=(x_i,y_i)$ and the center of the diatomic cell mapped by $\bm{R}$. The Hamiltonian is formulated in real-space and numerically diagonalized by solving the eigenvalue problem $\mathcal{H}\psi_n=E_n\psi_n$. The resulting eigenvalues $E_n$ and eigenstates $\psi_n$ serve as the basis for calculations of all quantities addressed in this work. The Hamiltonian is written as
\begin{equation}
\mathcal{H}=\mathcal{H}_{\rm SSH}+\mathcal{H}_{t'},
\end{equation}
where
\begin{equation}
\mathcal{H}_{\rm SSH}=
\sum_{\langle i,j\rangle}
\left(t-\tau_{ij}\delta t\right)
\left(c_i^\dagger c_j+\mathrm{H.c.}\right),
\label{eq:HSSH}
\end{equation}
with $\tau_{ij}=\pm1$, distinguishing inter- and intra-cell bonds. The condition $\delta t >0$ corresponds to the topological situation of the conventional SSH system \cite{SSH1979}. The staggered hopping term changes the strength of the dimerization in the model employing different values of $|\delta t|$. Moreover, the self-similar geometry of the Koch chain generates additional bonds connecting sites of the same sublattice located one bond length apart. These additional hoppings are addressed including 
\begin{equation}
\mathcal{H}_{t'}
=
t'
\sum_{\langle  i,j\rangle_{\mathrm{\triangle}}}
\left(c_i^\dagger c_j+\mathrm{H.c.}\right),
\label{eq:Htp}
\end{equation}
where the sum over $\langle i,j\rangle_{\mathrm{\triangle}}$ denotes pairs of sites connected at the triangular basis of the Koch fractal. For $t'=0$, the Hamiltonian satisfies the chiral symmetry $\{\Gamma,\mathcal{H}_{\rm SSH}\}=0$, with $\Gamma=
\sum_{i\in A}c_i^\dagger c_i-
\sum_{j\in B}c_j^\dagger c_j\,\,,$ being essentially a sub-lattice operator. Since $\mathcal{H}_{t'}$ couples identical sublattices, $\{\Gamma,\mathcal{H}\}\neq0$, and then $t'$ explicitly breaks the chiral symmetry [see Appendix~\ref{app:chiral_breaking}]. Therefore, the fractal geometry naturally provides a geometrical mechanism for chiral-symmetry breaking, absent in the conventional one-dimensional SSH chain.


Fig.~\ref{FIG_CHIRALKOCH}(a) illustrates the self-similar construction of the Koch--SSH chain. The triangles generated at successive iterations bring nonconsecutive sites into close spatial proximity, allowing additional hopping terms across the bases of the triangles. To preserve the alternating hopping sequence and suppress end-induced finite-size effects, we impose periodic boundary conditions by connecting sites $(N)$ and $(1)$. Since site $(N)$ belongs to sublattice $(B)$, and site $(1)$ belongs to sublattice $(A)$, the closing bond has amplitude $t+\delta t$, thus preserving the dimerization pattern. This boundary condition removes physical end states and allows the local marker to probe the bulk-like behavior of sufficiently large finite systems \cite{Melo2023,Weichen2023,Julia2022,Tiao2025,Oliveira2024}. In contrast to the nearest-neighbor SSH hoppings, which exclusively connect opposite sublattices, the additional $t'$ bonds connect sites belonging to the same sublattice. Consequently, these geometrically induced couplings explicitly break chiral, or sublattice, symmetry. Thus, the Koch geometry provides a deterministic microscopic mechanism for moving from the ideal chiral SSH limit, without including disorder or external perturbations.

The effect of this chiral-symmetry-breaking contribution on the half-filled spectral gap is shown in Fig.~\ref{FIG_CHIRALKOCH}(b). For $t'=0$, the system reduces to the conventional SSH chain. At $\delta t=0$, all nearest-neighbor hoppings are equal to $t$, and the resulting uniform chain is gapless. Thus, the gap closing at $\delta t=0$ marks the conventional SSH topological transition. A finite dimerization ($\delta t \neq0$) opens a gap according to $E_{\mathrm{gap}}/t=4|\delta t|$. Within our convention, $\delta t>0$ leads to larger values of $(t+\delta t)$ in comparison with the hopping $(t-\delta t)$, which corresponds to a nontrivial SSH phase, while $\delta t<0$ corresponds to the trivial phase \cite{SSH1979,Atala2013,Lu2024}.  

For $t'\neq0$, the gap closing at $\delta t=0$ is lifted and replaced by a finite minimum. In this situation, the spectral gap is no longer controlled exclusively by the SSH dimerization but results from the competition between $\delta t$ and the geometrically induced hopping $t'$. This competition also accounts for the intersections between the curves in Fig.~\ref{FIG_CHIRALKOCH}(b), where distinct combinations of $\delta t$ and $t'$ may yield the same half-filled gap, while their full spectra, eigenstates, and localization properties remain different. The corresponding energy spectra are presented in Figs.~\ref{FIG_CHIRALKOCH}(c)--(h). As expected, for $t'=0$, Figs.~\ref{FIG_CHIRALKOCH}(c)--(e) show references of the topological transitions that occur in the SSH model \cite{SSH1979,Atala2013,Lu2024}. The uniform chain in Fig.~\ref{FIG_CHIRALKOCH}(c), obtained for $\delta t=0$, displays the expected gap closing at half filling. For $\delta t/t=0.25$ and $=0.50$, shown in Fig.~\ref{FIG_CHIRALKOCH}(d) and Fig.~\ref{FIG_CHIRALKOCH}(e), respectively, the spectrum separates into particle-hole-symmetric sectors with gaps $E_{\mathrm{gap}}/t=1 \text{, and } =2$, consistent with the $E_{\mathrm{gap}}=4|\delta t|$ correspondence of . Because periodic boundary conditions are employed, these spectra describe the bulk levels and do not display the zero-energy end states expected for an open nontrivial SSH chain.

The corresponding Koch-connected systems at fixed $t'=t$ are shown in Figs.~\ref{FIG_CHIRALKOCH}(f)--(h). The energy spectrum for the isotropic Koch limit $(\delta t=0  \text{ and } t'=t)$ is illustrated in Fig.~\ref{FIG_CHIRALKOCH}(f), in which all allowed hopping amplitudes have the same magnitude. Despite the absence of SSH dimerization, the spectrum remains gapped at half filling because of the hierarchical B--B intra-sublattice connections. In addition, the spectrum becomes asymmetric around zero energy, indicating the break of particle-hole symmetry, and exhibits isolated clusters of states, separated by fractal-induced gaps, resulting in a characteristic staircase-like profile 
~\cite{Li2023,Jensen1983,LageSOC,Song2026}. Fig.~\ref{FIG_CHIRALKOCH} and Fig.(g)~\ref{FIG_CHIRALKOCH}(h) show the combined effects of this hierarchical connectivity and finite dimerization for $\delta t/t=0.25$ and $=0.50$, respectively. In this regime, increasing $\delta t$ modifies the separation between positive and negative solutions, and the gap no longer follows the simple SSH expression $4|\delta t|$. Instead, it possesses a non-linear behavior, as shown in Fig.~\ref{FIG_CHIRALKOCH}(b).

One interesting feature appearing in the isotropic Koch spectrum is the highly degenerate level at $E=-t$. Anticipating the discussion presented in Sec.~\ref{ap:minimal}, this degenerate level originates from destructive interference of states laying spatially around triangle sites, resulting in localized patterns along the triangular unit. Therefore, for $t'=t$, these states become degenerate at $E=-t$, giving rise to a flat spectral feature. Finite dimerization $(\delta t \neq 0 )$ makes the two sides of each triangular basis nonequivalent and consequently splits these localized states, as seen in panels Fig.~\ref{FIG_CHIRALKOCH}(g) and Fig.~\ref{FIG_CHIRALKOCH}(h) when compared with Fig.~\ref{FIG_CHIRALKOCH}(f) for $E=-t$. Thus, the Koch--SSH system supports a rich picture of spectral gaps caused by clustered energy levels and localized states whose spatial and topological properties can be further characterized using a local marker together with the inverse participation ratio, and local density of states. Moreover, because $t'$ breaks the protecting chiral symmetry, the conventional SSH winding number is no longer quantized. Thus, the real-space marker is essential for determining how the topological character evolves beyond the ideal chiral limit.

\begin{figure}[!h]
    \centering
\includegraphics[width=8cm]{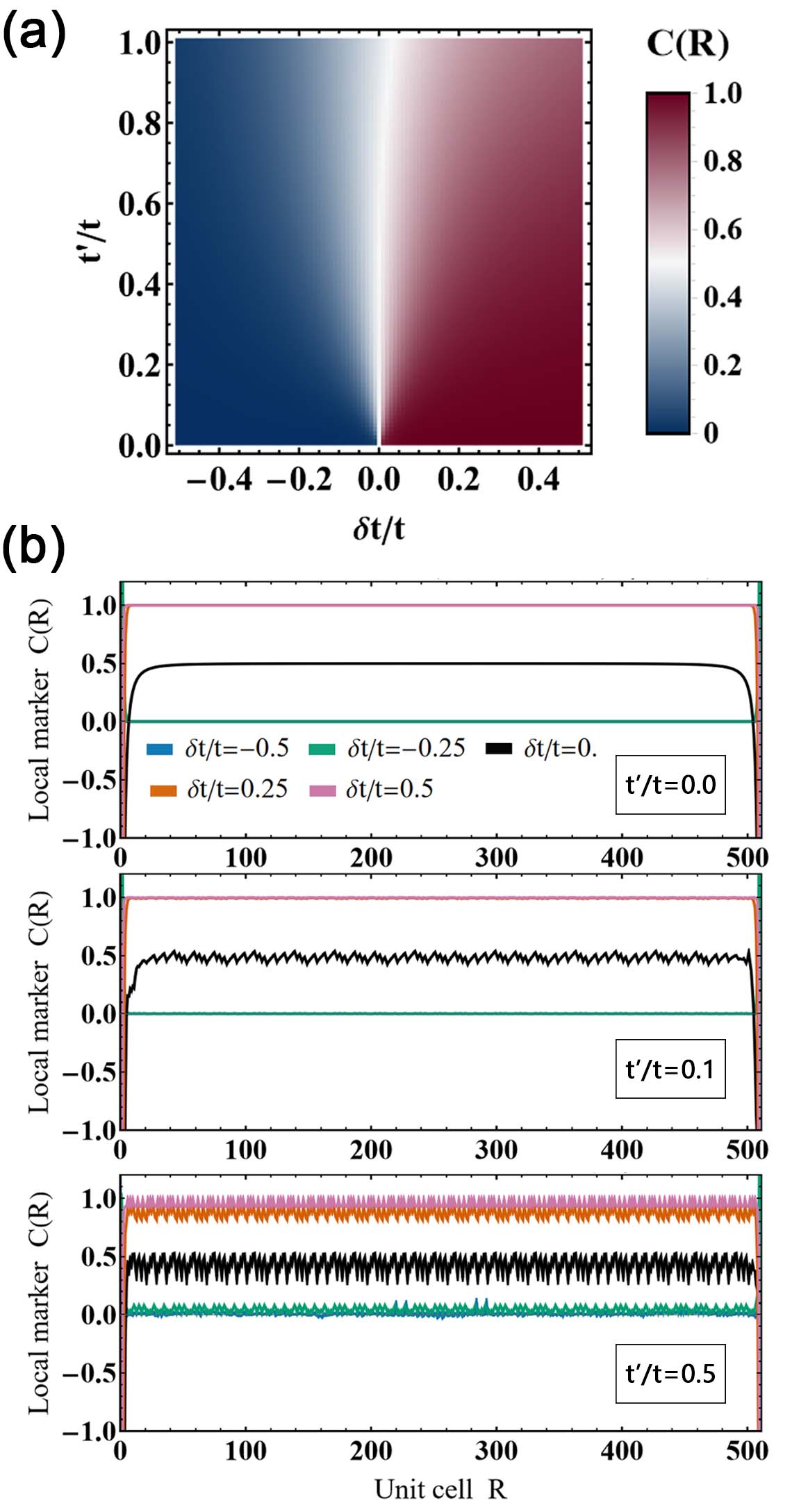}
   \caption{(a) Local marker parameter map for Koch--SSH G(5) with $C(R)$ evaluated over different $t'/t$ and $\delta t/t$ parameters for $\nu=1/2$. (b) Cross real-space resolution of the local marker for $\delta t/t =(-0.5,-0,25,0,0.25,0.5)$ and $t'/t=(0,0.1,0.5)$, marked with different colors.}
   \label{koch_diagram}
\end{figure}

\subsection{Local topological marker}

\begin{figure*}[t!]
\centering   
\includegraphics[width=16cm]{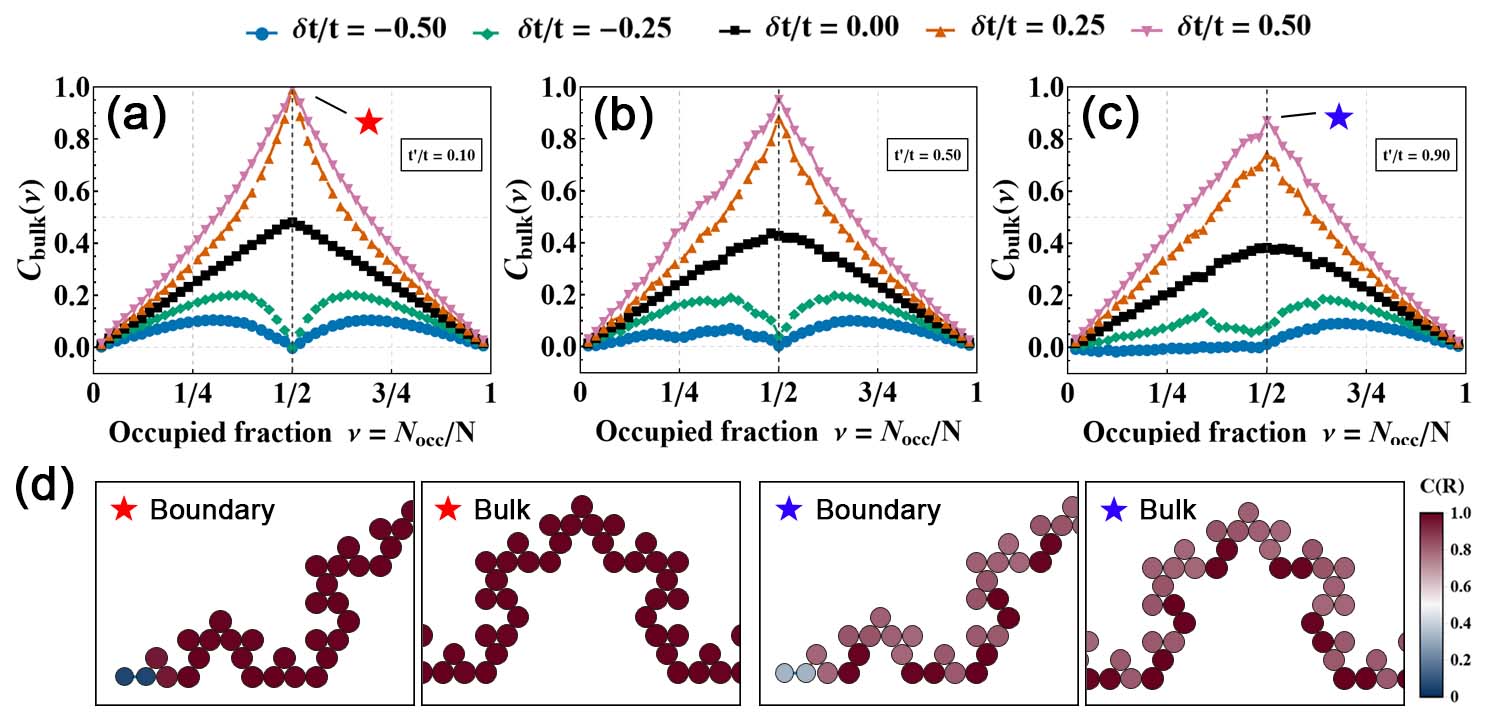}
   \caption{Local marker evaluated at the bulk region in function of the occupied fraction $\nu$, considering (a) $t'/t=0.1$ (b) $t'/t=0.5$, (c) $t'/t=0.9$ for different values of $\delta t$, given by distinct  colors. (d) Site-resolved resolution of $C(R)$ considering $\nu=1/2$ with $\delta t/t =0.0$ , and $\delta t=0.5$ (red and blue stars), with spatial zoom displacing boundary and bulk $C(R)$ resolutions.}
   \label{koch_marker2}
\end{figure*}
To characterize the topological properties of the Koch--SSH curve in real space, we employ the local topological marker introduced in Refs.~\cite{Bianco2011, Weichen2023}. Unlike the Zak phase, this quantity is defined locally and remains applicable in the absence of translational symmetry, making it particularly suitable for fractal lattices. The local marker operator is defined as 
\begin{equation}
\mathcal{C}=\Gamma \left(\mathcal{Q}x\mathcal{P}
+\mathcal{P}x\mathcal{Q}\right),
\label{eq:marker}
\end{equation}
where $x$ is the position operator along the chain,
and $\mathcal{P}$ and $\mathcal{Q}=\mathbb{1}-\mathcal{P}$ are the projectors onto the occupied and unoccupied single-particle states, respectively,
\begin{equation}
\mathcal{P}=\sum_{n=1}^{N_{occ}}
|\psi_n\rangle
\langle\psi_n|,
\qquad
    \mathcal{Q}=\sum_{n=N_{occ}+1}^{N}
|\psi_n\rangle
\langle\psi_n|.
\end{equation}
The occupation operator  is calculated at half-filling: $\nu=N_{\rm occ}/N=1/2$, except stated elsewhere. After diagonalizing the Hamiltonian, the marker is evaluated in real space from the diagonal elements of $\mathcal{C}$. Since each unit cell contains one $A$ and one $B$ site, the local marker is obtained as 
\begin{equation}
C(R)=
\langle A_R|\mathcal{C}|A_R\rangle
+
\langle B_R|\mathcal{C}|B_R\rangle,
\label{eq:localmarker}
\end{equation}
where $R$ labels the unit cells of the Koch chain. In the limit $t'=0$, Eq.~(\ref{eq:marker}) reduces to the local marker of the conventional SSH model, recovering the quantized topological value~\cite{Melo2023, Weichen2023,Oliveira2024}. Introducing the hopping $t'$ explicitly breaks chiral symmetry because of B--B intra-sublattice hoppings, allowing us to investigate how fractality modifies the spatial distribution of $C(R)$ and the robustness of the topological phase.

The topological transitions within the Koch--SSH model is summarized in Fig.~\ref{koch_diagram}(a), which presents the local marker diagram as a function of the dimerization $\delta t/t$ and the geometrically induced hopping $t'/t$. The color scale denotes the bulk local marker $C(R)$, averaged over the central region of the chain to avoid boundary condition effects. In the chiral-symmetric case ($t'=0$), the conventional SSH behavior is recovered with $C(R)=0$ for $\delta t < 0$ and $C(R) = 1$ for $\delta t>0$, capturing the trivial and non-trivial insulating regimes, respectively. The two phases are separated by an ill-defined region at $\delta t=0$, where the bulk gap closes and the local marker evaluated at the central region approaches $C(R)= 1/2$. At this situation, the system cross topological transition rather than forming a distinct half-quantized phase.

Increasing $t'$ values, connecting sites within the same B--B sublattice, break the chiral symmetry responsible for the quantization of $C(R)$,  removing the topological protection. The $C(R)$ spatial profiles in Fig.~\ref{koch_diagram}(b) show that $t'/t \neq 0$ promotes increasing oscillations in the local marker. The results show that the Koch geometry partially preserves the topological transition (trivial/nontrivial regimes) with quantized $C(R)$  for weak $t'$ hopping term, while stronger $t'$ drives the topological phase to nonquantized regions.




The spatially averaged bulk values of $C(R)$ are shown in Figs.~\ref{koch_marker2}(a)--(c)  as a function of the occupation fraction for different $t'/t$ ratios and $\delta t/t$ values (colored curves). At half filling ($\nu=1/2$), $C(R)$ remains close to unity for small $t'/t$ and $\delta t>0$, whereas for $\delta t <0$ it approaches $C(R)\approx 0$, as expected for a trivial phase. At $\delta t=0.5$, the local marker approaches $C(R)=0.5$ values, signalizing the gapless transition in the SSH model, where the topological marker is no longer quantized. As $t'/t$ increases, the asymmetry of the curves becomes increasingly pronounced, reflecting the chiral-symmetry breaking induced by the intra-sublattice hopping.

Fig.~\ref{koch_marker2}(d) shows the spatially resolved $C(R)$ at $\nu=1/2$, for $\delta t/t=0.5$. Red and blue stars indicate the values of the $C(R)$ evaluated at representative sites, for $t/'t=0.1$ and $t'/t=0.9$, respectively, located at the boundary, and within bulk regions of the Koch fractal. For $t'/ t=0.9$ (blue star), $C(R)=1$ is still obtained at a few sites located between regions where $C(R)$ is non-quantized. In contrast, for $t'/t =0.1$  (red star) the vast majority of sites exhibit $C(R)=1$, indicating a substantially more robust topological character throughout the bulk region.  

\section{Isotropic Koch limit}\label{sec:isotropic_koch}

\begin{figure*}[t!]
\centering   
\includegraphics[width=18cm]{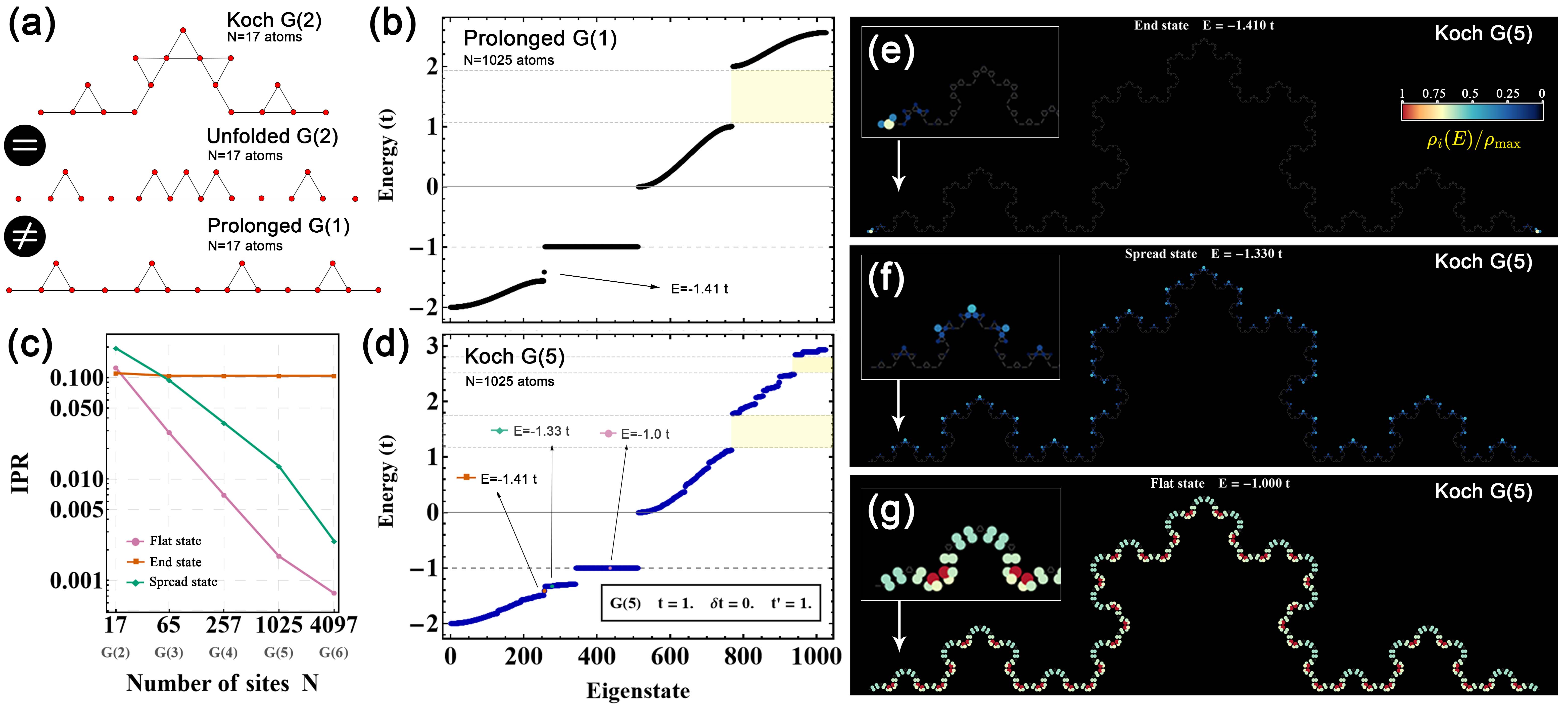}
\caption{Electronic structure and localization in the isotropic ($t'=t$, $\delta t=0$) Koch chain. (a) Koch $G(2)$, its unfolded representation, and a prolonged nonfractal chain of repeated $G(1)$ units. (b) Spectrum of the prolonged finite chain with $N=1025$, showing the flat state at $E=-t$ and an end state near $E=-1.41t$. (c) IPR calculated at $E=-t$. End state, and spread state near $E=-1.33t$ for different Koch generations. (d) Spectrum of the Koch $G(5)$ chain; shaded regions highlight one example of the fractal-induced gaps within the hierarchical staircase structure. (e)--(g) Real-space LDOS of the end, spread, and flat state, respectively, for $E=-1.41t,-1.33t,-t$, normalized by a common $\rho_{\max}$.}
\label{FIGKOCH_FULLPAGE}
\end{figure*}

We now investigate the electronic consequences of fractal connectivity in the isotropic limit of the Koch--SSH model. This regime is obtained by setting $t=t'$, and $\delta t=0$. The dimerization is removed, and all bonds of the Koch curve carry the same hopping amplitude. The Hamiltonian reduces to
\begin{equation}
\mathcal{H}_{\mathrm{K}}
=
t\sum_{\langle i,j\rangle_{\mathrm{K}}}
\left(
c_i^{\dagger}c_j+\mathrm{H.c.}
\right),
\label{eq:H_Koch_isotropic}
\end{equation}
where $\langle i,j\rangle_{\mathrm{K}}$ denotes pairs of sites connected by an edge of the complete Koch network. This limit eliminates the SSH hopping imbalance and isolates the spectral features generated exclusively by the hierarchical fractal geometry. To isolate the effects of the hierarchical connectivity of the Koch curve, we compare the related geometries shown in Fig.~\ref{FIGKOCH_FULLPAGE}(a). The first corresponds to the genuine $G(2)$ Koch fractal with open boundary conditions. We then consider an unfolded representation of this structure, in which the sites are arranged linearly while the sequence and connectivity of the local triangular units are preserved, given by two isolated $G(1)$ triangles in each side, and three triangles connected in the middle. For comparison, we also construct an extended nonfractal chain by periodically repeating the $G(1)$ triangular unit, namely as prolonged $G(1)$. As illustrated in Fig.~\ref{FIGKOCH_FULLPAGE}(a), the unfolded $G(2)$ geometry is not equivalent to a simple repetition of $G(1)$ units. Although we always consider the same number of atoms for the two structures, only the unfolded situation corresponds to the hierarchical connectivity from the Koch fractal.  
Fig.~\ref{FIGKOCH_FULLPAGE}(b) and Fig.~\ref{FIGKOCH_FULLPAGE}(d) compares the energy spectra for both the prolonged chain $G(1)$ and the Koch fractal $G(5)$, with the same number of atoms $N=1025$, where the number of atoms in the $n$-th Koch curve is given by $N=4^n+1$ sites \cite{Macia1998}. We notice that the systems have a highly degenerate state at $E_{\rm flat}=-t$. This common feature originates from the local triangular connectivity and is not, by itself, a consequence of fractality. In Sec.~\ref{ap:minimal} we investigate the origin of this flat state in a four-site analog, where the destructive interference explains the pattern observed in the finite counterparts. The finite energy spectrum in Fig.~\ref{FIGKOCH_FULLPAGE}(b) has four energy clusters; $E\geq2t$, $1\geq E/t \geq 0$, $E=-t$, and $-2\leq E/t \leq -1.5$, while a single solution, namely end state, appears at $E_{\rm end}=-1.41t \approx-\sqrt{2}t$ due to the open boundary conditions. Similarly, the Koch G(5) spectrum in Fig.~\ref{FIGKOCH_FULLPAGE}(d) also has the end state at the same energy level, while the clusters of solutions are separated into subsets, marked, for instance, by the yellow regions, where gaps are introduced, forming a staircase profile. The split of the energy levels seen for the Koch solution is a consequence of induced fractal gaps that occur due to the Koch connection-hierarchy. At each generation, replicated structures are coupled through a restricted set of intermediary sites that separate the group of interconnected consecutive triangles [as seen in the three middle-triangles of unfolded $G(2)$ in Fig.~\ref{FIGKOCH_FULLPAGE}(a)] with the isolated nonconsecutive triangles, e.g., the ones at the spatial edges of the unfolded $G(2)$ [see  Fig.~\ref{FIGKOCH_FULLPAGE}(a)]. This structure is absent from the prolonged $G(1)$ chain, which keeps the isolated triangles but lacks the hierarchical connectivity of the Koch construction. In fact, when the fractality is imposed, the state at $E=-t$ is less degenerate, decreasing from $255\times$ to $170\times$, for prolonged $G(1)$ and Koch $G(5)$, respectively. This happens because in the Koch curve, part of the degenerate states is being redistributed in energy due to the self-similarity of the system, and in the limit of $n\rightarrow\infty$ the degeneracy factor converges to a fixed scaling of the spectral weight as will be discussed in Sec~\ref{app:fractal_induced}. Additional states appear between $E=-1.41t$ and $E=-t$  in the fractal spectrum. We analyze further the state $E=-1.33t$, namely here as a spread state.

In fractal structures the localization properties can be classified adopting the inverse participation ratio (IPR) \cite{TitusSC2018,PaiPrem2019,Manna2024}. Here we analyze the IPR of highlighted states in Koch $G(5)$ employing,
\begin{equation}
\mathrm{IPR}_{n}
=
\frac{\sum_i |\psi_n(i)|^4}
{\left(\sum_i|\psi_n(i)|^2\right)^2} \ .
\label{eq:IPR_Koch}
\end{equation}
The IPR is plotted as a function of the original number of sites $N$, with both axes represented on logarithmic scales. As shown in Fig.~\ref{FIGKOCH_FULLPAGE}(c), the IPR of the end-state ($E=-1.41t$) approaches a finite value as the generation increases, demonstrating that its spatial distribution remains confined to a finite number of sites [see Fig.~\ref{FIGKOCH_FULLPAGE}(e)] at the edges of the Koch curve. In contrast, the decreasing IPR of the state near $E=-1.33t$ indicates a progressive increase in its spatial spreading, consistent with the spread profile displayed in Fig.~\ref{FIGKOCH_FULLPAGE}(f). The resulting local density of states (LDOS) extends through the chain while being strongly concentrated in specific spatial regions [see Fig.~\ref{FIGKOCH_FULLPAGE}(g)]. The IPR results reflects the increasing spatial localized contributions along the chain, rather than a uniform delocalization (IPR $\propto N^{-1}$). More generally, fractal systems can exhibit unconventional scaling of the form $\mathrm{IPR}\propto N^{\gamma}$ \cite{PhysRevB.110.035403,Manna2024}. 


The normalized LDOS maps in Figs.~\ref{FIGKOCH_FULLPAGE}(e--g) further illustrate the spatial characteristics associated with these distinct regimes. At $E=-t$, the spectral weight is concentrated on repeated triangular patterns distributed along the Koch structure. The state near $E=-1.41t$ (end state) is instead totally confined to the two open boundaries, with maximum $|\psi|^2$ weight close to the end sites. Finally, the spread state at $E=-1.33t$ is distributed over the complete structure but follows the self-similar arrangement of the Koch curve, producing a triangular pattern rather than a homogeneous bulk distribution. The three LDOS maps are normalized using a common maximum, allowing their relative spectral weights to be compared directly.
The results reveal few distinct geometrical mechanisms, where the local triangular loops generate the highly degenerate level at $E=-t$ through destructive interference. The hierarchical connectivity of the Koch curve produces a staircase-like spectral structure and spatial modulation of the extended states. The end state originates from the open terminations. By comparing Fig.~\ref{FIG_CHIRALKOCH}(f) with Fig.~\ref{FIGKOCH_FULLPAGE}(d) we see that the end states disappear when the two outermost sites are connected to form a closed geometry \cite{lagefrontier}. Although the LDOS obtained by summing all degenerate contributions at $E=-t$ extends over the entire Koch structure, this does not imply that each eigenstate in the degenerate level is spatially extended. Instead, it can be represented by combinations of compact localized modes protected by the triangular geometry alone. The LDOS at $E=-t$ in Fig.~\ref{FIGKOCH_FULLPAGE}(g) further reveals a selective localization pattern, where the largest spectral weights occur on isolated triangles separated from neighboring sites or groups of triangles introduced by the Koch hierarchy. Groups of consecutive connected triangles exhibit vanishing contributions at this energy [i.e. the three triangles in the middle of Koch $G(2)$ in Fig.~\ref{FIGKOCH_FULLPAGE}(a)]. This behavior reflects additional destructive-interference constraints imposed by the hierarchical connectivity. In the following sections, we examine the origin of this localization pattern and demonstrate the emergence of compact localized states in the corresponding periodic triangular analogy.


\section{Minimal triangular model for the flat state}
\label{ap:minimal}

\begin{figure}[!h]
    \centering
\includegraphics[width=6.5cm]{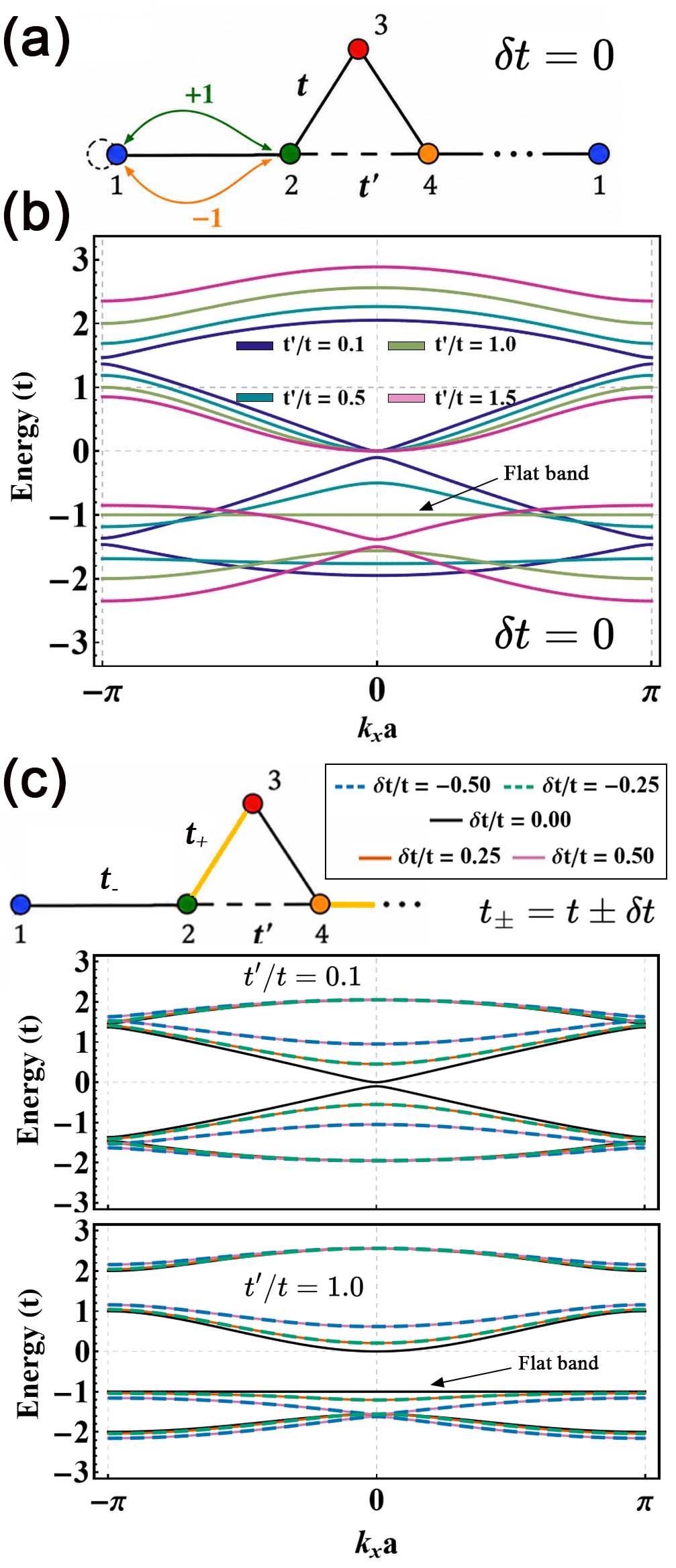}
   \caption{(a) Illustration of the repeating unit cell and the interference pattern at site 1 in $E=-t$ for $t=t'$. (b) Electronic bands of the one-dimensional system comparing different $t'/t$ values. (c) Band evolution for considering the dimerization with different colors for each $\delta t $ value and $t_{\pm}=t\pm\delta t$, for $t'/t=0.1$ and $t'/t=0.9$, respectively, on top and bottom panels.}
   \label{FIG:MINIMALa}
\end{figure}

Previous studies have investigated one-dimensional chains composed of triangular geometries, in which compact localized states  emerge through destructive interferences, leading to robust spatially localized states~\cite{Leykam2017,Maimaiti2019,Kempkes2023}. Here, we want to clarify the origin and robustness of the highly degenerate states that have appeared at $E=-t$ by varying the parameters $t',$ and $t$. For that, we consider a minimal one-dimensional periodic model containing four sites per unit cell. This reduced system preserves the local triangular connectivity of the isotropic Koch structure and is suitable for describing the prolonged $G(1)$ situation. We investigate the implications of breaking isotropy by setting $t'\neq t$ and different dimerization values $\delta t /t$. We first consider the isotropic case $t'=t$ and $\delta t/t=0$, which provides a reference point for identifying the microscopic mechanism responsible for the flat state independently of the global fractal geometry. In the Bloch basis $\Psi=\left(
\psi_1,\psi_2,\psi_3,\psi_4
\right)^{\top}$,
the Hamiltonian is written as,
\begin{equation}
H(k_x) 
=
\begin{pmatrix}
0 & 1 & 0 & e^{-ik_x a} \\
1 & 0 & 1 & t'/t\\
0 & 1 & 0 & 1 \\
e^{ik_xa} & t'/t & 1 & 0 
\end{pmatrix} \ ,
\label{eq:minimal_Hk}
\end{equation}
The intracell hoppings describe the path
$1-4$ together with the additional bond
$2-4$, which closes the triangle [see Fig.~\ref{FIG:MINIMALa}(a)]. The phase factor $e^{-ik_xa}$ originates from the hopping between site $1$ and site $4$ of adjacent unit cells. The associated eigenvector is 
\begin{equation}
    |\psi(k_x)\rangle \propto \begin{pmatrix}
        1-t'/t  \\
        t'/t - 1 - e^{-i k_x a} \\
        e^{-i k_x a} - t'/t \\
        1 
        
    \end{pmatrix} .
\end{equation}
The destructive interference required to generate the flat state demands an isotropic condition in which all bonds have the same hopping amplitude ($t'=t$). Solving the secular equation for this case, one gets
\begin{eqnarray}
(E+t)   \left[E^{3} -tE^{2}  -4t^{2}E+2t^{3}-2t^{3}\cos k_xa\right]=0.
\label{eq:minimal_characteristic}
\end{eqnarray}
The first factor gives the dispersionless band,
$E_{\mathrm{flat}}=-t,$
which is independent of the crystal momentum. 
The remaining three bands are dispersive.
The existence of the flat band in this simple periodic model demonstrates that the $E=-t$ states do not require the complete Koch hierarchy. Instead, it originates locally from the triangular hopping geometry shared by both the prolonged and fractal structures analyzed previously. The eigenvector associated with the flat band is given by 
\begin{equation}
\left|\psi_{\mathrm{flat}}(k_x)\right\rangle
=
\frac{1}{\sqrt{4-2\cos k_xa}}
\begin{pmatrix}
0\\
-e^{-ik_x a}\\
e^{-ik_x a}-1\\
1
\end{pmatrix}.
\label{eq:minimal_flat_eigenvector}
\end{equation}
As the wave function amplitude at the first site is zero, $\psi_1(k)=0$, the amplitudes on sites $2$ and $4$ satisfy the relation  $\psi_2(k)+e^{-ik_x a}\psi_4(k)=0$. Consequently, the flat state cannot propagate through the bond connecting neighboring unit cells. Thus, the absence of dispersion is produced by destructive quantum interference on the triangular geometry, forming compact localized states \cite{Leykam2017,Maimaiti2019}. At the Brillouin zone boundary $k_x =\pi/a$, the flat-band eigenvector becomes $\left|\psi_{\mathrm{flat}}(\pi/a)\right\rangle
=
\sqrt{1/6}(
0,
1,
-2,
1)^{\top}   $
with probability distribution
$\left|\psi_{\mathrm{flat}}(\pi/a)\right|^{2}
=
\left(
0,{1}/{6},{2}/{3},{1}/{6}
\right)^{\top}.$
The wave probability is then distributed on the three sites forming the triangle, with no weight on the site coupling consecutive cells. This compact localization explains why repeated triangular geometries generate a large number of states at the same energy $E=-t$. Fig.~\ref{FIG:MINIMALa}(b) shows the comparison between the bands as $t'/t$ is changed while $\delta t/t=0$. We notice that the flat state only occurs for the condition $t'/t=1.$ 

We next examine the effect of hopping dimerization over the range $-0.5\leq\delta t/t\leq0.5$. Fig.~\ref{FIG:MINIMALa}(c) shows the corresponding hopping pattern, in which the alternating bonds are $t_{\pm}=t\pm\delta t$, while the additional hopping $t'$ connects sites $2$ and $4$ and closes the triangle. The upper and lower graphs correspond to bands calculated for $t'/t=0.1$ and $t'/t=1.0$, respectively. In the conventional SSH limit ($t'=0$) the bipartite Hamiltonian possesses chiral symmetry, which enforces the electron-hole symmetry and produces a gap closing at $\delta t=0$. A finite $t'$ connects sites belonging to the same sublattice and breaks this symmetry. Consequently, already for $t'/t=0.1$, the spectrum becomes slightly asymmetric about zero energy and the SSH gap closing is replaced by a small gap, leaving the system gapped through the displayed dimerization range. The spectral asymmetry becomes more pronounced as $t'/t$ increases. A distinct feature appears in the lower panel for the simultaneous conditions $t'/t=1$ and $\delta t=0$, represented by the black curve, where one band becomes completely dispersionless at $E=-t$. Considering, Fig.~\ref{FIG:MINIMALa}(b) and Fig.~\ref{FIG:MINIMALa}(c), both show that exact flat states requires hopping isotropy and the absence of dimerization to be satisfied. When $t'/t=1$ but $\delta t\neq0$, the unequal triangle-side hoppings $t_+$ and $t_-$ introduce a finite dispersion. Conversely, setting $\delta t=0$ still not generates the flat state when $t'/t\neq1$, since the triangle-base hopping remains different from the side hoppings. Exact flat states therefore requires both $t'=t$ and $\delta t=0$, for which all three hoppings around the triangle are equal. This hopping-isotropic configuration supports the destructive-interference condition responsible for the flat band and degenerated states within the finite models described in Sec.~\ref{sec:isotropic_koch}.


\section{Degeneracy and spectral weight of the flat state}\label{app:fractal_induced}

To distinguish the effects of the local triangular geometry from those produced by the self-similar Koch connectivity, we compare the Koch ($K$) curve $G(n)$ with the prolonged ($P$) isotropic ($t'=t,\delta t=0$) finite chains. Both geometries have exactly the same number of sites, $N=4^{n}+1$, where $n$ is indexed by the generations of the Koch curve. Therefore, the comparison is performed between Hamiltonians of the same dimension, under the same open boundary conditions, and with the same hopping amplitude. Any spectral difference between them can consequently be attributed to their distinct connectivity rather than to a finite-size mismatch. Although both systems contain the local triangular geometry responsible for destructive interference, they differ in how this geometry is distributed along the chain.

As previously discussed, the prolonged structure contains one triangle per repeated $G(1)$ unit, giving
$N_{\triangle}^\mathrm{P}=4^{n-1}.$ By contrast, the iterative construction of the Koch curve generates triangular structures at different hierarchical connections, leading to $N_{\triangle}^{\mathrm{K}}
=({4^{n}-1})/{3}.$ Since, in that case, a regular chain contains $(4^n)$ bonds, the total number of bonds for each system is $N_b^{\mathrm{P}} =5\times4^{n-1}$, and $N_b^{\mathrm{K}} 
=({4^{n+1}-1})/{3}$. Computing the difference between the number of bonds ($\Delta N_b $) and triangles ($\Delta N_{\triangle}$) in each case gives the same result:
\begin{equation}
N_b^{\mathrm{K}}
-N_b^{\mathrm{P}}
=
N_{\triangle}^{\mathrm{K}}
-N_{\triangle}^{\mathrm{P}}
=
\frac{4^{n-1}-1}{3}. \label{eq:excessoN}
\end{equation}
The additional structural features of the Koch geometry give rise to higher-coordination connections that are not present in the prolonged chain. Therefore, the Koch curve contains more triangles than the prolonged structure; however, these triangles are not independent.  Instead, they overlap through the hierarchical connectivity, imposing additional constraints and quantum interferences on the wave-function amplitudes.

\begin{figure}[!h]
    \centering
\includegraphics[width=8.5cm]{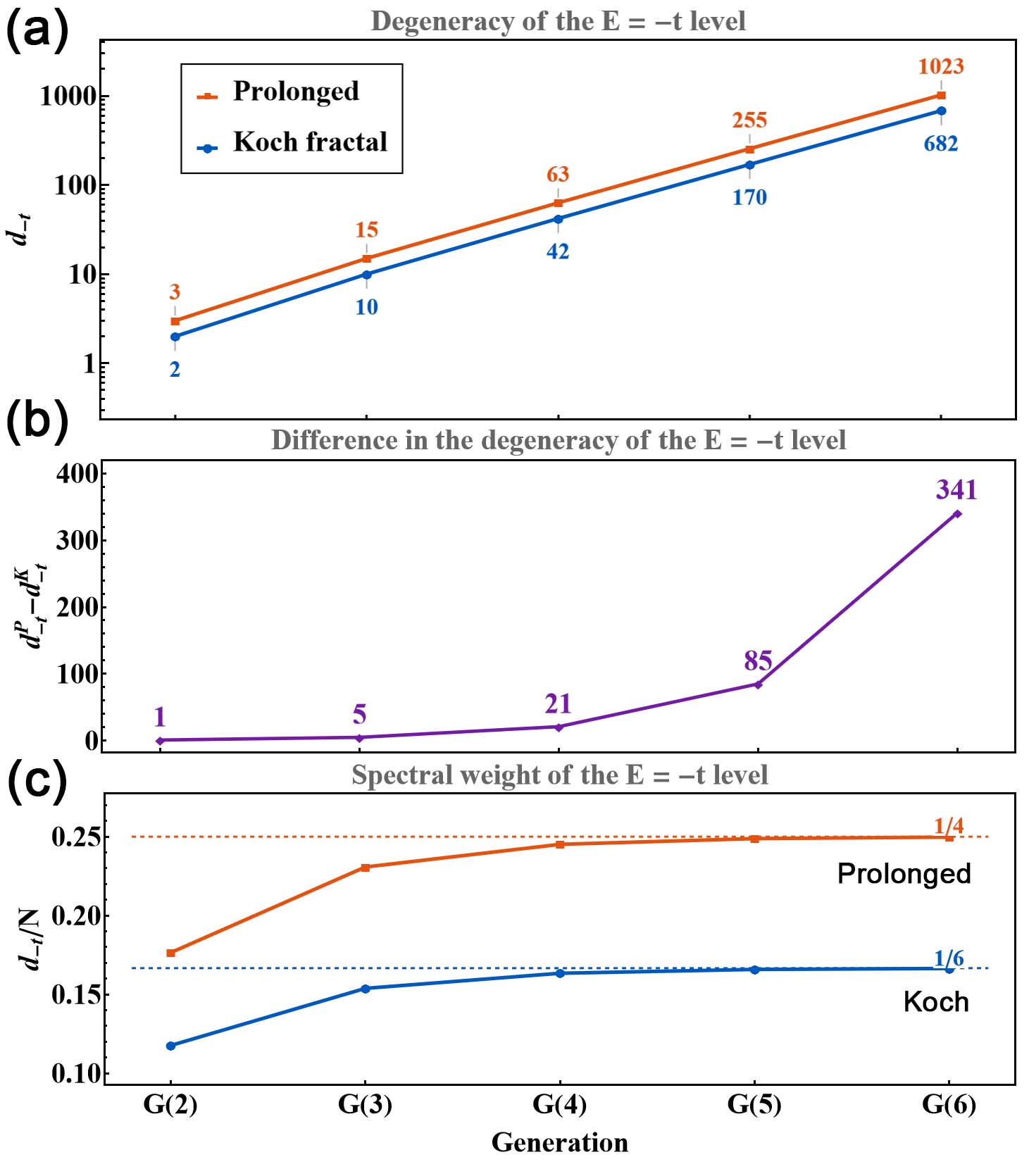}
   \caption{Degeneracy analysis for the flat state $E=-t$ along different generations of the Koch with equivalent number of sites in the prolonged version (a) Logarithmic scale for each degeneracy in orange and blue, for prolonged and Koch (b) Difference of the degeneracy values shown in panel (a). (c) Degeneracy normalized by the number of atoms in both cases, where the generations .}
   \label{FIG:ap1:FIG1}
\end{figure}
A physical consequence of this distinction is particularly clear for the $E=-t$ state shown in Fig.~\ref {FIG:ap1:FIG1}. As discussed for the periodic analogue, the presence of this energy level in both geometries confirms that it is fixed by destructive interference associated with the local triangular geometry. Repeating this geometry produces a family of states for which propagation through the structure is suppressed, generating an extensively degenerate flat-band that satisfies the destructive-interference conditions. Fig.~\ref {FIG:ap1:FIG1}(a) shows the evolution of the $E=-t$ degeneracy on a logarithmic scale. The degeneracy at $E=-t$ for the prolonged system, assumes the numerical values $3,15,63,255,$ and $1023$, obtained from $G(2)$ to $G(6)$, that follow the rule
\begin{equation}
d_{-t}^{\mathrm{P}} 
=4^{n-1}-1.
\end{equation}
Notice that it grows with the number of triangles minus one. For the Koch curve, the corresponding sequence $2,10,42,170,$ and $682$ is described by
\begin{equation}
d_{-t}^{\mathrm{K}} 
=
\frac{2}{3}\left(4^{n-1}-1\right).
\end{equation}
Both degeneracies grow with the same exponential factor, showing that the $E=-t$ state remains present in the two structures. Nevertheless, the Koch degeneracy is systematically reduced according to
\begin{equation}
d_{-t}^{\mathrm{K}} 
=
\frac{2}{3}
d_{-t}^{\mathrm{P}} .
\end{equation}
Thus, the hierarchical connectivity removes one third of the states that would otherwise belong to the $E=-t$ state of the prolonged geometry. This reduction is illustrated in Fig.~\ref {FIG:ap1:FIG1}(b), where
\begin{equation}
\Delta d_{-t} 
=
d_{-t}^{\mathrm{P}} 
-d_{-t}^{\mathrm{K}} 
\
=
\frac{4^{n-1}-1}{3}.
\end{equation}
Remarkably, $\Delta d_{-t} $ is identical to both the difference in the number of triangles and the difference in the number of bonds in the Koch structure in Eq.~\ref{eq:excessoN}, i.e., 
$\Delta d_{-t}=\Delta N_{\triangle}=\Delta N_b$.  This equality demonstrates that the reduction in the degeneracy at $E=-t$ is directly associated with the additional connections introduced by the Koch hierarchy. Each additional triangle requires an extra closing bond, which removes one independent state from the $E=-t$ state compared with the prolonged structure. Consequently, when consecutive triangles occur along the Koch curve, their destructive-interference conditions are no longer independent, leading to a lower degeneracy than that of the prolonged chain. This behavior can be clearly observed by revisiting Fig.~\ref{FIGKOCH_FULLPAGE}(g), where the localization pattern follows the isolated, nonconsecutive triangles separated by an intermediary site. In contrast, the consecutive interconnected triangles of the Koch geometry do not contribute to charge localization at $E=-t$. Their absence at this energy directly accounts for the reduced degeneracy. Thus, although the Koch curve contains a larger number of triangles, its hierarchical connections promote the overlap of localized modes, thereby reducing the number of linearly independent states at $E=-t$. The distinction survives in the large-generation limit, as demonstrated by the spectral weights in Fig.~\ref{FIG:ap1:FIG1}(c). For the prolonged system,
\begin{equation}
\frac{d_{-t}^{\mathrm{P}} }{N}
=
\frac{4^{n-1}-1}{4^n+1}
\xrightarrow[n\rightarrow\infty]{}
\frac{1}{4},
\end{equation}
where for the Koch curve,
\begin{equation}
\frac{d_{-t}^{\mathrm{K}} }{N}
=
\frac{2\left(4^{n-1}-1\right)}
{3\left(4^n+1\right)}
\xrightarrow[n\rightarrow\infty]{}
\frac{1}{6}.
\end{equation}
The nonzero limiting values demonstrate that $E=-t$ level remains macroscopically degenerate in both systems and is not a boundary-induced or accidental finite-size feature. However, the fraction of the spectrum carried by this level is reduced from $1/4$ in the prolonged chain to $1/6$ in the Koch fractal. Thus, the comparison separates two complementary physical mechanisms; the local triangular geometry pins the common energy $E=-t$ through destructive interference, while the hierarchical Koch connectivity determines the dimension and spectral weight of the resulting degenerate states.



\section{Conclusions}
We investigated how fractal connectivity modifies symmetry-protected topology and localization in an SSH model defined on the Koch curve. The triangular geometry introduces B--B intra-sublattice hoppings $t'$, breaking the chiral symmetry. In the chiral-symmetric limit, the real-space local marker identifies the quantized trivial and nontrivial SSH phases separated by the gap closing at $\delta t=0$. For finite $t'$, the spectral symmetry and topological marker quantization are lost,  characterized by noninteger and spatially oscillating marker values.

In the isotropic limit, $t'=t$ and $\delta t=0$, the local triangular geometry and the global Koch hierarchy play distinct roles. Destructive interference within the triangles generates the highly degenerate flat level at $E=-t$, which is reproduced analytically by a periodic four-site model. Its elementary compact localized states extend over the triangular units and have zero amplitude on the connecting sites. Although the summed spectral weight of the degenerate level can cover the entire structure, it represents the collective projector onto all such states rather than a single extended eigenstate. The Koch hierarchy does not create this flat level. Instead, it reduces its degeneracy by approximately one third relative to the prolonged $G(1)$ chain, and reorganizes the spectrum into a staircase profile. The IPR quantity further distinguishes spread states, whose values decrease with system size, from the state near $E=-1.41t$ characterized by a finite IPR value. Overall, B--B intra-sublattice hopping at the bottom of the triangles leads to chiral-symmetry breaking. Moreover, the triangular geometry produces compact flat-band states, where the self-similar hierarchy governs their degeneracy and spectral intricate distribution. These findings establish fractal unique structures as prominent systems to explore topological transitions caused by geometrical constraints imposed by the self-similar pattern. 

The proposed model is also directly relevant to wave-based realizations. In photonic systems, for instance, the chain sites may be represented by coupled waveguides, with the hopping amplitudes controlled by the separation and overlap between neighboring modes \cite{BlancoRedondo2016,Fedorova2019,Savelev2020,Chen2021Waveguides}. The alternating SSH couplings and the additional triangle-base coupling $t'$ can be engineered geometrically, allowing the chiral-symmetry-breaking transition, the flat band at $E=-t$, and the associated interference patterns to be probed through optical intensity distributions and propagation dynamics \cite{Xia2018,CaceresAravena2022,Song2026}. These possibilities make Koch-based photonic and acoustic lattices promising platforms for experimentally exploring topology and localization generated purely by self-similar geometry.


\begin{acknowledgments} 

This work received partial support from the National Council for Scientific and Technological Development (CNPq). L.L.L. acknowledges financial support from the National Institute of Science and Technology Nanocarbon and 2D Materials (INCT Nanocarbono e Materiais 2D). A.L. thanks the financial support of FAPERJ (Grant No. E-26/200.569/2023). The authors acknowledge Pedro B. Melo and Rodrigo Arouca for the fruitful discussions.

\end{acknowledgments}

\appendix 
\setcounter{figure}{0}
\renewcommand{\thefigure}{A\arabic{figure}}

\section{Localized states in the minimal model}\label{app:LDOS}

\begin{figure}[!h]
    \centering
\includegraphics[width=8.5cm]{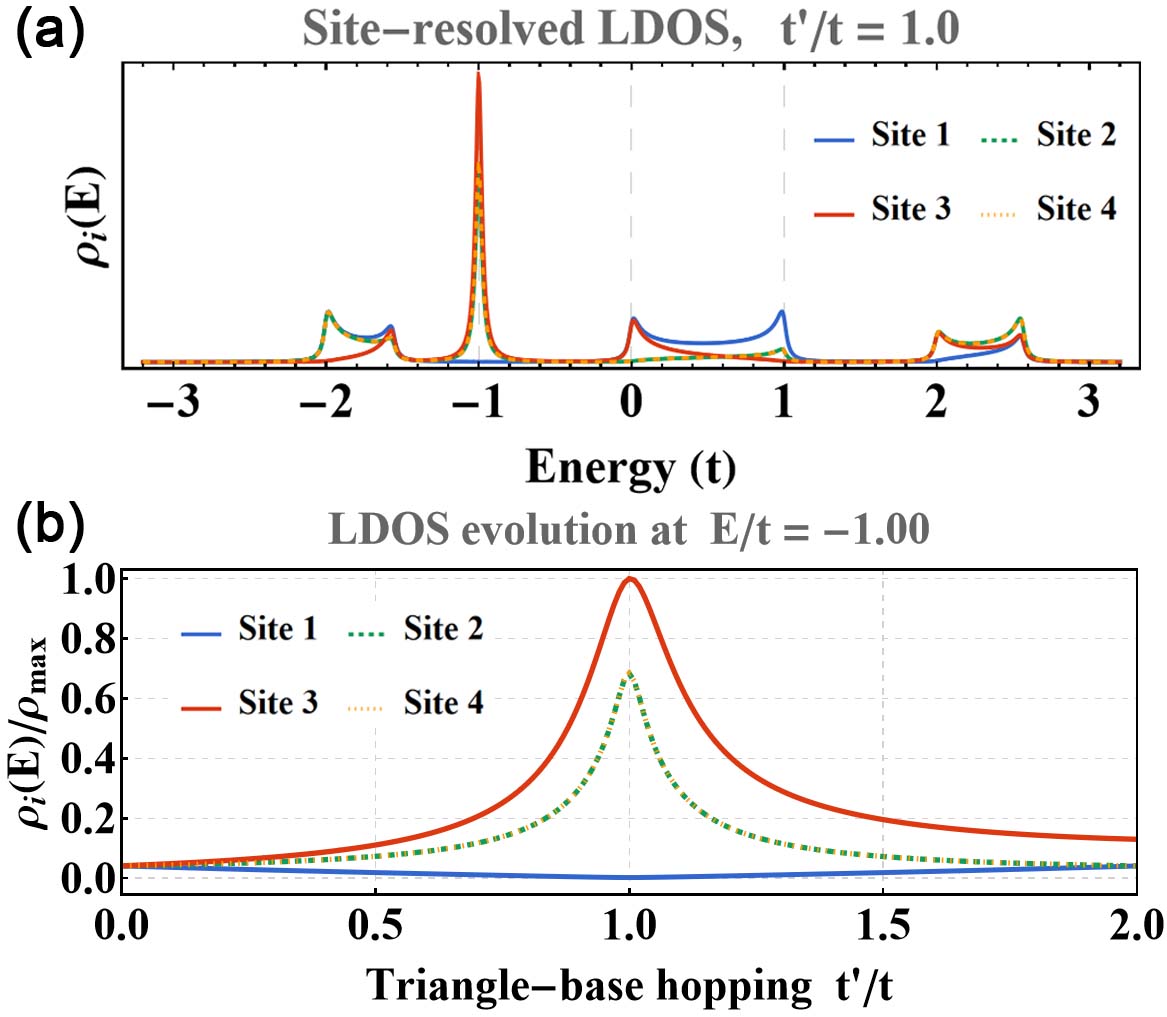}
   \caption{ (a) Site-resolved LDOS for $t'=t$ over the whole energy spectra with different colors representing the calculation over distinct sites. Sites 2 and 4 contributes equally, while the biggest contribution at $E=-t$ is due to sites 2, 3, and 4. (b) LDOS evolution at $E=-t$ normalized by the max. localized energy state for different ratios of $t'/t$.}
   \label{FIG:MINIMALb}
\end{figure}

Fig.~\ref{FIG:MINIMALb}(a) shows the site-resolved local density of states $\rho_i(E)$ (LDOS) of the four-site model calculated with
\begin{equation}
\rho_i(E)
= 
\sum_{k,n}
\left|\psi_{n,i}(k)\right|^2
\frac{\eta/\pi}
{\left[E-E_n(k)\right]^2+\eta^2},
\label{eq:ldos}
\end{equation}
where $\eta=0.025t$ is a Lorentzian broadening parameter. The sharp peak at $E=-t$ is the contribution of the flat band. The probability density of this energy
is distributed over the three sites forming the triangle; sites $2$ and $4$ contribute
equally, while site $3$ carries the largest weight of $|\psi|^2$. Otherwise, the site-resolved LDOS vanishes at site $1$. This distribution
follows directly from the flat-band eigenvector, for which
$\psi_1(k)=0$ and $|\psi_2(k)|=|\psi_4(k)|$. The suppression of the LDOS on site $1$ is the real-space signature of destructive interference, which prevents the flat-band state from propagating
between neighboring unit cells.

Fig.~\ref{FIG:MINIMALb}(b) shows the LDOS at the fixed energy
$E=-t$ as a function of the hopping ratio $t'/t$. The $|\psi|^2$ weight is maximal at the isotropic point $t'/t=1$, where the destructive-interference condition produces an exactly dispersionless band pinned at $E=-t$. Moving $t'$ away from $t$ values removes this exact condition, causing
the band to become dispersive or shift away from $E=-t$ and consequently reducing the LDOS at the chosen energy. The triangle sites dominate the spectral response, with
$\rho_2(-t)=\rho_4(-t)$ by symmetry, where the site responsible for the contribution between neighboring cell connections reaches its minimum at the isotropic point. Thus, the resonance at $t'/t=1$ identifies the geometric condition under which the triangular unit hosts the compact flat-band state. Excepting the trivial situation of $t'/t=0$, the site 3 contribution overcomes the other sites in the complete analysis. 


\section{Chiral-symmetry breaking in the Koch--SSH model}
\label{app:chiral_breaking}

In the sublattice basis, the SSH Hamiltonian, the chiral operator \cite{Asboth2016,Chiu2016}, and
the additional B--B intra-sublattice hopping term can be written as
\begin{equation}
h_{\rm SSH}=
\begin{pmatrix}
0&D\\ D^\dagger&0
\end{pmatrix},
\
\Gamma=
\begin{pmatrix}
I_A&0\\0&-I_B
\end{pmatrix},
\
h_{t'}=
\begin{pmatrix}
T_A&0\\0&T_B
\end{pmatrix}, \nonumber
\label{eq:Hssh_appendix}
\end{equation}
where $D$ contains the hoppings between opposite sublattices, with
$T_A$ and $T_B$ describing the triangle-base hoppings $t'$ between sites
of the same sublattice. Since $\Gamma^\dagger=\Gamma$ and
$\Gamma^2=I$, these contributions transform as
\begin{equation}
\Gamma h_{\rm SSH}\Gamma=-h_{\rm SSH},
\qquad
\Gamma h_{t'}\Gamma=h_{t'}.
\label{eq:chiral_transformations}
\end{equation}
Thus, $h_{\rm SSH}$ anticommutes with $\Gamma$ and has a spectrum
symmetric about zero energy, while $h_{t'}$ commutes with $\Gamma$. This structure can be illustrated using the closed four-site Koch
$G(1)$ unit introduced in Sec.~\ref{ap:minimal}. In the path ordering
$(1,2,3,4)=(A_1,B_1,A_2,B_2)$, with
$t_\pm=t\pm\delta t$, the complete Hamiltonian is
\begin{equation}
h^{G(1)}
=
\begin{pmatrix}
0&t_-&0&t_+\\
t_-&0&t_+&t'\\
0&t_+&0&t_-\\
t_+&t'&t_-&0
\end{pmatrix}.
\label{eq:Hfull_G1_sitebasis}
\end{equation}
The hopping $t'$ connects sites $2$ and $4$, both belonging to
sublattice $B$. Reordering the basis as
$(1,3,2,4)=(A_1,A_2,B_1,B_2)$ gives
\begin{eqnarray}
h^{G(1)}
&=&
\begin{pmatrix}
0_{2\times2}&D_{G(1)}\\
D_{G(1)}^\dagger&T_B^{G(1)} 
\end{pmatrix},
\end{eqnarray}
\begin{eqnarray}
D_{G(1)}&=&
\begin{pmatrix}
t_-&t_+\\
t_+&t_- 
\end{pmatrix},\
\end{eqnarray}
\begin{eqnarray}
T_B^{G(1)}=
\begin{pmatrix}
0&t'\\
t'&0

\end{pmatrix}. 
\end{eqnarray}
The diagonal sublattice block $T_B^{G(1)}$ makes the breaking
of chiral symmetry explicit. More generally, for
$h=h_{\rm SSH}+h_{t'}$,
\begin{equation}
\Gamma h\Gamma=-h_{\rm SSH}+h_{t'} 
\rightarrow
\{\Gamma,h\}
=2\Gamma h_{t'}
=2
\begin{pmatrix}
T_A&0\\
0&-T_B
\end{pmatrix}. \nonumber
\label{eq:anticommutator}
\end{equation}
Consequently, the full Koch--SSH Hamiltonian anticommutes with
$\Gamma$ only when the B--B intra-sublattice blocks vanish. In the present
construction, the symmetry breaking is controlled directly by $t'$,
and the conventional chiral-symmetric SSH limit is recovered as
$t'\rightarrow0$.


\bibliography{refs}

\end{document}